Mathematica Italia
User Group Meeting

# Investigating indipendent subsets of graphs, with Mathematica

Pietro Codara, Ottavio D'Antona

Dipartimento di Informatica, Università degli Studi di Milano

## Abstract

With this work we aim to show how Mathematica can be a useful tool to investigate properties of combinatorial structures. Specifically, we will face enumeration problems on independent subsets of powers of paths and cycles, trying to highlight the correspondence with other combinatorial objects with the same cardinality. Then we will study the structures obtained by ordering properly independent subsets of paths and cycles. We will approach some enumeration problems on the resulting partially ordered sets, putting in evidence the correspondences with structures known as Fibonacci and Lucas Cubes.

## 1. Introduction

This work follows step by step the research carried out in [CD12a]. The focus, however, is different. While [CD12a] provide results supported by analytical proofs, as in a classical mathematical paper, here we want to show in what stages of our research work, and how, *Mathematica* can help us. Much of the research that follows is purely enumerative. We will see how, in each case of counting, it is useful to have a *Mathematica* implementation of our formulas and to observe the tabulated results. In particular, when we are able to conjecture two or more different formulas to perform the same calculation, *Mathematica* can give us immediate feedback about the quality of our hypothesis. This will be valid, in particular, at the end of our paper, when, thanks to *Mathematica*, we can make assumptions for a theorem which still, at the moment, does not have any analytical proof. For all unexplained notions and for the full proofs of mathematical results contained in this piece of work, please see [CD12a, CD12b].

For a graph $G$ we denote by $V(G)$ the set of its vertices, and by $E(G)$ the set of its edges.

### Definition 1.1.

For $n, h \geq 0$,

- the *h-power of a path*, denoted by $P_n^{(h)}$, is a graph with $n$ vertices $v_1, v_2, \ldots, v_n$ such that, for $1 \leq i, j \leq n, i \neq j, (v_i, v_j) \in E(P_n^{(h)})$ if and only if $|j-i| \leq h$;
- the *h-power of a cycle*, denoted by $Q_n^{(h)}$, is a graph with $n$ vertices $v_1, v_2, \ldots, v_n$ such that, for $1 \leq i, j \leq n, i \neq j, (v_i, v_j) \in E(Q_n^{(h)})$ if and only if $|j-i| \leq h$ or $|j-i| \geq n-h$.

Thus, for instance, $P_n^{(0)}$ and $Q_n^{(0)}$ are the graphs made of $n$ isolated nodes, $P_n^{(1)}$ is the path with $n$ vertices, and $Q_n^{(1)}$ is the cycle with $n$ vertices. Next figure shows the powers of paths $P_n^{(h)}$, for $n = 1, \ldots, 6$.

Out[75]=

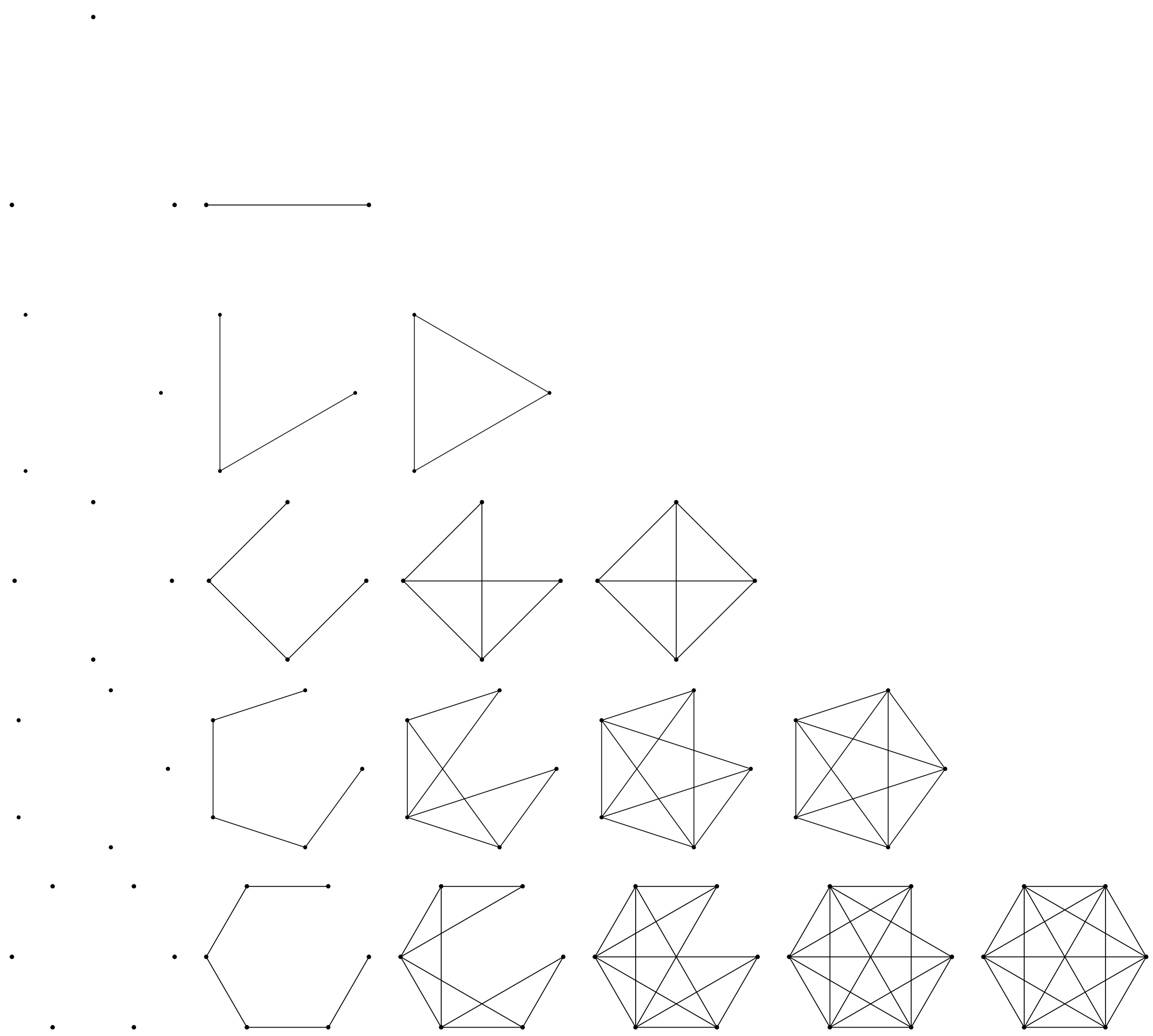

Next figure shows the powers of cycles $Q_n^{(h)}$, for $n = 3, \ldots, 6$.

Out[82]=

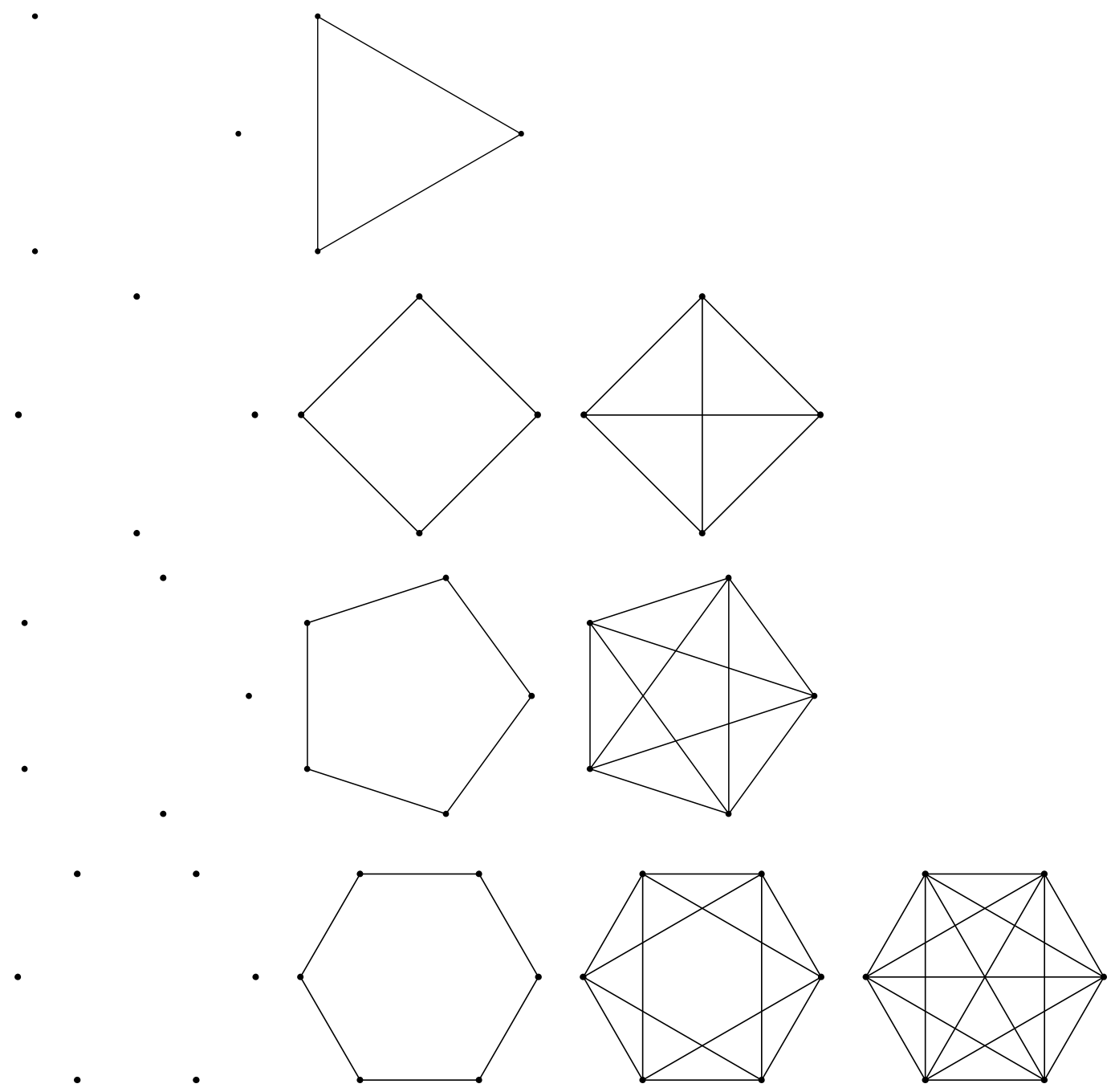

## Definition 1.2.

An *independent subset of a graph* G is a subset of $V(G)$ not containing adjacent vertices.

Let $H_n^{(h)}$, and $M_n^{(h)}$ be the Hasse diagrams of the posets of independent subsets of $P_n^{(h)}$, and $Q_n^{(h)}$, respectively, ordered by inclusion. Clearly, $H_n^{(0)} \cong M_n^{(0)}$ is a Boolean lattice with $n$ atoms ($n$-cube, for short).

Following [MS02] (see also [Kla11]), a *Fibonacci string of order n* is a binary strings of length $n$ without (two) consecutive 1's. Recalling that the Hamming distance between two binary strings $\alpha$ and $\beta$ is the number $H(\alpha, \beta)$ of bits where $\alpha$ and $\beta$ differ, we can define the *Fibonacci cube of order n*, denoted $\Gamma_n$, as the graph $(V, E)$, where $V$ is the set of all Fibonacci strings of order $n$ and, for all $\alpha, \beta \in V$, $(\alpha, \beta) \in E$ if and only if $H(\alpha, \beta) = 1$. Fibonacci cubes were introduced as an interconnection scheme for multicomputers in [Hsu93], and their combinatorial structure has been further investigated, *e.g.* in [KP07, MS02]. Several generalizations of the notion of Fibonacci cubes has been proposed (see, *e.g.*, [IKR12a, Kla11]).

A *Lucas cube of order n*, denoted $\Lambda_n$, is defined as the graph whose vertices are the binary strings of length $n$ without either two consecutive 1's or a 1 in the first and in the last position, and in which the vertices are adjacent when their Hamming distance is exactly 1 (see [MPCZS01]). A generalization of the notion of Lucas cubes has been proposed in [IKR12b].

In the first part of the paper we evaluate $p_n^{(h)}$, *i.e.*, the number of *independent subsets* of $P_n^{(h)}$, and $H_n^{(h)}$, *i.e.*, the number of edges of $H_n^{(h)}$.  In the second part of the paper we derive similar results for $q_n^{(h)}$, *i.e.*, the number of *independent subsets* of $Q_n^{(h)}$, and $M_n^{(h)}$, *i.e.*, the number of edges of $M_n^{(h)}$. We will see, moreover, that $H_n^{(h)}$ and $M_n^{(h)}$ are another generalizations (new, as far as we know) of the notions of Fibonacci and Lucas cubes, respectively. For each result found, we will implement the corresponding formulas in *Mathematica* and show a table of results.

# 2. The independent subsets of powers of paths

For $n, h, k \geq 0$, we denote by $p_{n,k}^{(h)}$ the number of independent $k$-subsets of $P_n^{(h)}$.

## Lemma 2.1.

For $n, h, k \geq 0$,

$$p_{n,k}^{(h)} = \binom{n - h\,k + h}{k}.$$

This result is Theorem 1 of [Hog70]. With the help of *Mathematica*, we can immediately obtain some value of $p_{n,k}^{(h)}$, for $h = 1, \ldots, 4$.

```
MyBinomial[n_, k_] := If[n < k, 0, Binomial[n, k]]
```

```
p[h_, n_, k_] := MyBinomial[n - h k + h, k]
```

### $p_{n,k}{}^{(1)}$

Out[97]//TableForm=

| | n=0 | 1 | 2 | 3 | 4 | 5 | 6 | 7 | 8 | 9 | 10 | 11 | 12 | 13 |
|---|---|---|---|---|---|---|---|---|---|---|---|---|---|---|
| k=0 | 1 | 1 | 1 | 1 | 1 | 1 | 1 | 1 | 1 | 1 | 1 | 1 | 1 | 1 |
| 1 | 0 | 1 | 2 | 3 | 4 | 5 | 6 | 7 | 8 | 9 | 10 | 11 | 12 | 13 |
| 2 | 0 | 0 | 0 | 1 | 3 | 6 | 10 | 15 | 21 | 28 | 36 | 45 | 55 | 66 |
| 3 | 0 | 0 | 0 | 0 | 0 | 1 | 4 | 10 | 20 | 35 | 56 | 84 | 120 | 165 |
| 4 | 0 | 0 | 0 | 0 | 0 | 0 | 0 | 1 | 5 | 15 | 35 | 70 | 126 | 210 |
| 5 | 0 | 0 | 0 | 0 | 0 | 0 | 0 | 0 | 0 | 1 | 6 | 21 | 56 | 126 |
| 6 | 0 | 0 | 0 | 0 | 0 | 0 | 0 | 0 | 0 | 0 | 0 | 1 | 7 | 28 |
| 7 | 0 | 0 | 0 | 0 | 0 | 0 | 0 | 0 | 0 | 0 | 0 | 0 | 0 | 1 |
| 8 | 0 | 0 | 0 | 0 | 0 | 0 | 0 | 0 | 0 | 0 | 0 | 0 | 0 | 0 |

### $p_{n,k}{}^{(2)}$

Out[98]//TableForm=

| | n=0 | 1 | 2 | 3 | 4 | 5 | 6 | 7 | 8 | 9 | 10 | 11 | 12 | 13 |
|---|---|---|---|---|---|---|---|---|---|---|---|---|---|---|
| k=0 | 1 | 1 | 1 | 1 | 1 | 1 | 1 | 1 | 1 | 1 | 1 | 1 | 1 | 1 |
| 1 | 0 | 1 | 2 | 3 | 4 | 5 | 6 | 7 | 8 | 9 | 10 | 11 | 12 | 13 |
| 2 | 0 | 0 | 0 | 0 | 1 | 3 | 6 | 10 | 15 | 21 | 28 | 36 | 45 | 55 |
| 3 | 0 | 0 | 0 | 0 | 0 | 0 | 0 | 1 | 4 | 10 | 20 | 35 | 56 | 84 |
| 4 | 0 | 0 | 0 | 0 | 0 | 0 | 0 | 0 | 0 | 0 | 1 | 5 | 15 | 35 |
| 5 | 0 | 0 | 0 | 0 | 0 | 0 | 0 | 0 | 0 | 0 | 0 | 0 | 0 | 1 |
| 6 | 0 | 0 | 0 | 0 | 0 | 0 | 0 | 0 | 0 | 0 | 0 | 0 | 0 | 0 |

### $p_{n,k}{}^{(3)}$

Out[99]//TableForm=

| | n=0 | 1 | 2 | 3 | 4 | 5 | 6 | 7 | 8 | 9 | 10 | 11 | 12 | 13 | 14 |
|---|---|---|---|---|---|---|---|---|---|---|---|---|---|---|---|
| k=0 | 1 | 1 | 1 | 1 | 1 | 1 | 1 | 1 | 1 | 1 | 1 | 1 | 1 | 1 | 1 |
| 1 | 0 | 1 | 2 | 3 | 4 | 5 | 6 | 7 | 8 | 9 | 10 | 11 | 12 | 13 | 14 |
| 2 | 0 | 0 | 0 | 0 | 0 | 1 | 3 | 6 | 10 | 15 | 21 | 28 | 36 | 45 | 5 |
| 3 | 0 | 0 | 0 | 0 | 0 | 0 | 0 | 0 | 0 | 1 | 4 | 10 | 20 | 35 | 5 |
| 4 | 0 | 0 | 0 | 0 | 0 | 0 | 0 | 0 | 0 | 0 | 0 | 0 | 0 | 1 | 5 |
| 5 | 0 | 0 | 0 | 0 | 0 | 0 | 0 | 0 | 0 | 0 | 0 | 0 | 0 | 0 | 0 |

## $p_{n,k}{}^{(4)}$

Out[100]//TableForm=

| | n=0 | 1 | 2 | 3 | 4 | 5 | 6 | 7 | 8 | 9 | 10 | 11 | 12 | 13 | 14 |
|---|---|---|---|---|---|---|---|---|---|---|---|---|---|---|---|
| k=0 | 1 | 1 | 1 | 1 | 1 | 1 | 1 | 1 | 1 | 1 | 1 | 1 | 1 | 1 | 1 |
| 1 | 0 | 1 | 2 | 3 | 4 | 5 | 6 | 7 | 8 | 9 | 10 | 11 | 12 | 13 | 14 |
| 2 | 0 | 0 | 0 | 0 | 0 | 0 | 1 | 3 | 6 | 10 | 15 | 21 | 28 | 36 | 45 |
| 3 | 0 | 0 | 0 | 0 | 0 | 0 | 0 | 0 | 0 | 0 | 0 | 1 | 4 | 10 | 20 |
| 4 | 0 | 0 | 0 | 0 | 0 | 0 | 0 | 0 | 0 | 0 | 0 | 0 | 0 | 0 | 0 |

The coefficients $p_{n,k}^{(h)}$ also enjoy the following property: $p_{n,k}^{(h)} = p_{n-k+1,k}^{(h-1)}$.

For $n$, $h \geq 0$, the number of all independent subsets of $P_n^{(h)}$ is

$$p_n^{(h)} = \sum_{k\geq 0} p_{n,k}^{(h)} = \sum_{k=0}^{\lceil n/(h+1)\rceil} p_{n,k}^{(h)} = \sum_{k=0}^{\lceil n/(h+1)\rceil} \binom{n-h\,k+h}{k}.$$

Some values of $p_n^{(h)}$ are shown in the following table, obtained by implementing our formula in *Mathematica*.

```
p[h_, n_] := Sum[p[h, n, k], {k, 0, Ceiling[n/(h + 1)]}]
```

Out[108]//TableForm=

| | n=0 | 1 | 2 | 3 | 4 | 5 | 6 | 7 | 8 | 9 | 10 | 11 | 1 |
|---|---|---|---|---|---|---|---|---|---|---|---|---|---|
| h=0 | 1 | 2 | 4 | 8 | 16 | 32 | 64 | 128 | 256 | 512 | 1024 | 2048 | 40 |
| 1 | 1 | 2 | 3 | 5 | 8 | 13 | 21 | 34 | 55 | 89 | 144 | 233 | 3 |
| 2 | 1 | 2 | 3 | 4 | 6 | 9 | 13 | 19 | 28 | 41 | 60 | 88 | 1 |
| 3 | 1 | 2 | 3 | 4 | 5 | 7 | 10 | 14 | 19 | 26 | 36 | 50 | 6 |
| 4 | 1 | 2 | 3 | 4 | 5 | 6 | 8 | 11 | 15 | 20 | 26 | 34 | 4 |
| 5 | 1 | 2 | 3 | 4 | 5 | 6 | 7 | 9 | 12 | 16 | 21 | 27 | 3 |
| 6 | 1 | 2 | 3 | 4 | 5 | 6 | 7 | 8 | 10 | 13 | 17 | 22 | 2 |
| 7 | 1 | 2 | 3 | 4 | 5 | 6 | 7 | 8 | 9 | 11 | 14 | 18 | 2 |
| 8 | 1 | 2 | 3 | 4 | 5 | 6 | 7 | 8 | 9 | 10 | 12 | 15 | 1 |
| 9 | 1 | 2 | 3 | 4 | 5 | 6 | 7 | 8 | 9 | 10 | 11 | 13 | 1 |
| 10 | 1 | 2 | 3 | 4 | 5 | 6 | 7 | 8 | 9 | 10 | 11 | 12 | 1 |

## Remark.

Denote by $F_n$ the $n^{th}$ element of the Fibonacci sequence $F_1 = 1$, $F_2 = 1$, and $F_i = F_{i-1} + F_{i-2}$, for $i > 2$ (the first values of the Fibonacci sequence are shown below). Then, $p_n^{(1)} = F_{n+2}$ is the number of elements of the Fibonacci cube of order $n$.

| n=1 | 2 | 3 | 4 | 5 | 6 | 7 | 8 | 9 | 10 | 11 | 12 | 13 |
|---|---|---|---|---|---|---|---|---|---|---|---|---|
| 1 | 1 | 2 | 3 | 5 | 8 | 13 | 21 | 34 | 55 | 89 | 144 | 233 |

Statement like the one in the remark can be also checked in Mathematica by displaying a table showing the positions on which two, or more, different formulas agree.

```
TableForm[Table[p[1, n] == MyFibo[n + 2], {n, 0, 17}],
 TableHeadings → {{"n=0", "1", "2", "3", "4", "5", "6", "7",
    "8", "9", "10", "11", "12", "13", "14", "15", "16", "17"}},
 TableAlignments → Center, TableDirections → Row]
```

| n=0 | 1 | 2 | 3 | 4 | 5 | 6 | 7 | 8 | 9 | 10 | 11 | 12 | 13 | 14 | 1! |
|---|---|---|---|---|---|---|---|---|---|---|---|---|---|---|---|
| True | True | True | True | True | True | True | True | True | True | True | True | True | True | True | Tı |

The following, simple fact is crucial for our work.

## Lemma 2.2.

For $n$, $h \geq 0$,

$$p_n^{(h)} = \begin{cases} n+1 & \text{if } n \leq h+1\,, \\ p_{n-1}^{(h)} + p_{n-h-1}^{(h)} & \text{if } n > h+1\,. \end{cases}$$

The following table, displaying some values of $p_n^{(h)}$ computed via the formula provided in Lemma 2.2, constitute a preliminary check on the identity between this formula and the one introduced before.

```
p2[h_, n_] := If[n > h + 1, p[h, n - 1] + p[h, n - h - 1], n + 1]
```

| | n=0 | 1 | 2 | 3 | 4 | 5 | 6 | 7 | 8 | 9 | 10 | 11 | 1 |
|---|---|---|---|---|---|---|---|---|---|---|---|---|---|
| h=0 | 1 | 2 | 4 | 8 | 16 | 32 | 64 | 128 | 256 | 512 | 1024 | 2048 | 40 |
| 1 | 1 | 2 | 3 | 5 | 8 | 13 | 21 | 34 | 55 | 89 | 144 | 233 | 3 |
| 2 | 1 | 2 | 3 | 4 | 6 | 9 | 13 | 19 | 28 | 41 | 60 | 88 | 1 |
| 3 | 1 | 2 | 3 | 4 | 5 | 7 | 10 | 14 | 19 | 26 | 36 | 50 | 6 |
| 4 | 1 | 2 | 3 | 4 | 5 | 6 | 8 | 11 | 15 | 20 | 26 | 34 | 4 |
| 5 | 1 | 2 | 3 | 4 | 5 | 6 | 7 | 9 | 12 | 16 | 21 | 27 | 3 |
| 6 | 1 | 2 | 3 | 4 | 5 | 6 | 7 | 8 | 10 | 13 | 17 | 22 | 2 |
| 7 | 1 | 2 | 3 | 4 | 5 | 6 | 7 | 8 | 9 | 11 | 14 | 18 | 2 |
| 8 | 1 | 2 | 3 | 4 | 5 | 6 | 7 | 8 | 9 | 10 | 12 | 15 | 1 |
| 9 | 1 | 2 | 3 | 4 | 5 | 6 | 7 | 8 | 9 | 10 | 11 | 13 | 1 |
| 10 | 1 | 2 | 3 | 4 | 5 | 6 | 7 | 8 | 9 | 10 | 11 | 12 | 1 |

# 3. The poset of independent subsets of powers of paths

The following picture shows the Hasse diagrams $H_n^{(h)}$ for $n = 1, \ldots, 4$, and $h \geq 0$. (Whenever $h \geq n$, $H_n^{(h)}$ coincide with $H_{h-1}^{(h)}$, and it is not shown).

Out[219]=

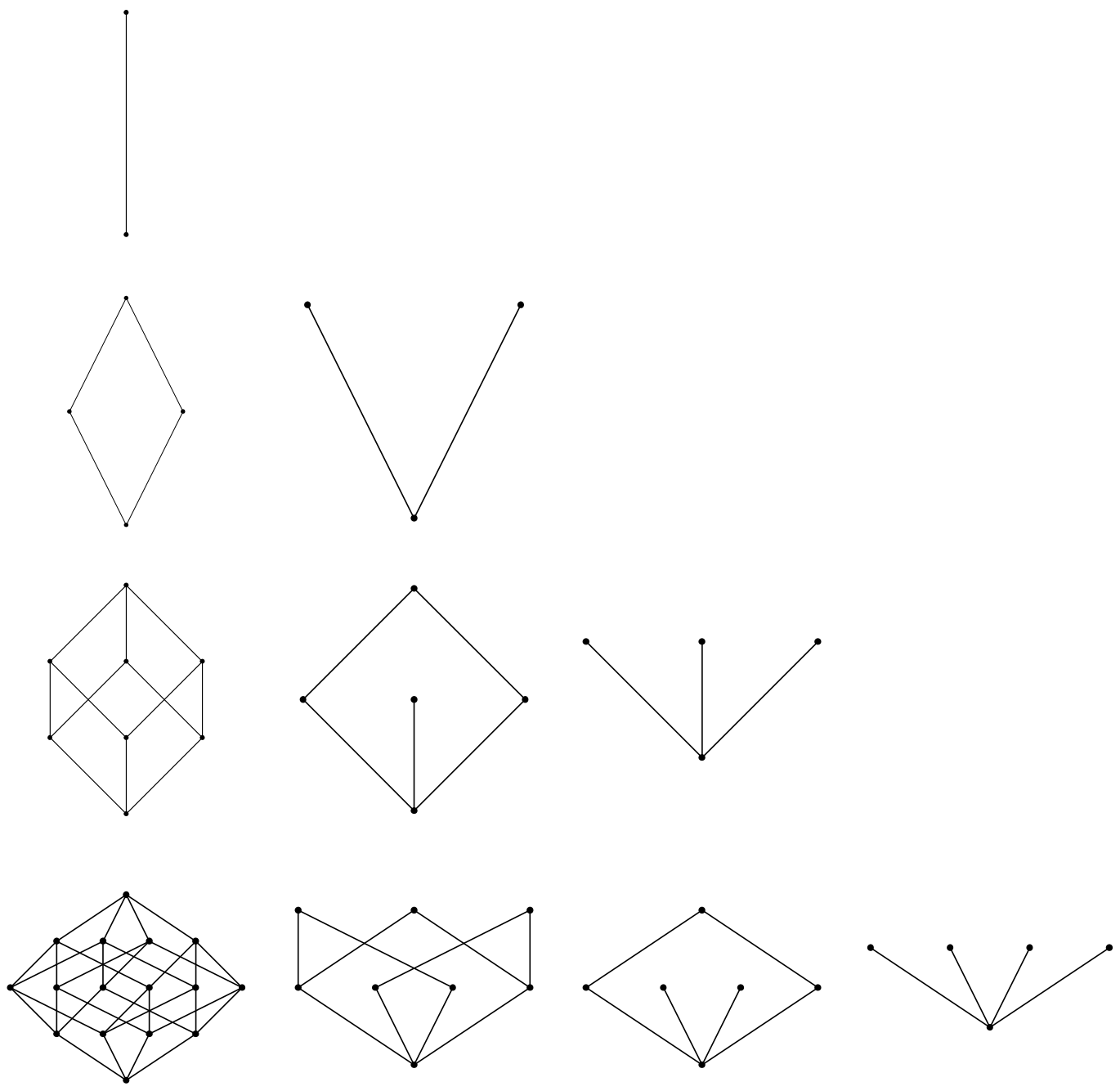

Notice that, as stated in the introduction, the first column - $H_n^{(0)}$ - is formed by the $n$-cubes, and the second column - $H_n^{(1)}$ - is the column of the Fibonacci cubes $\Gamma_n$. To check the latter, Fibonacci cubes can also be drawn by mean of their definition in terms of Fibonacci strings.

```
FibonacciCube[n_] := HasseDiagram[MakeGraph[FibonacciStrings[n],
    HammingDistance[#1, #2] == 1 && Count[#1, 1] < Count[#2, 1] &]];
```

In[247]:=
```
ShowGraphArray[Table[FibonacciCube[n], {n, 1, 5}]]
```

Out[247]=

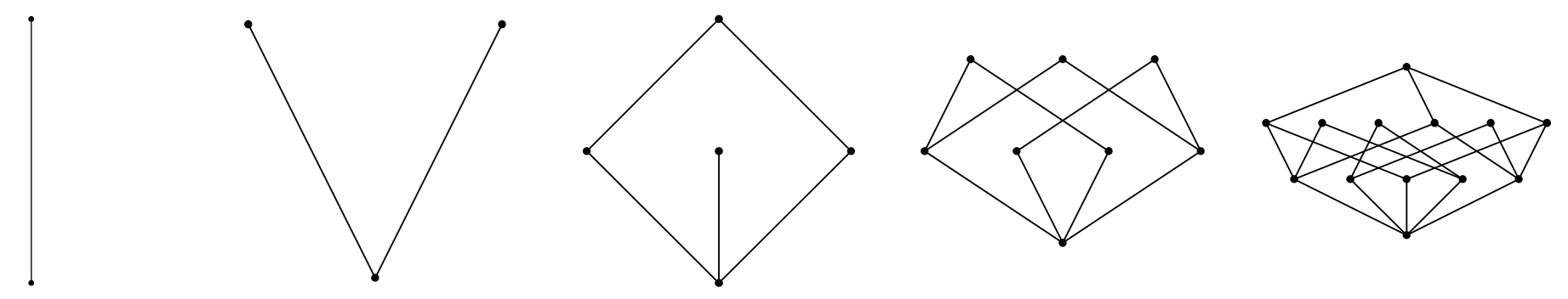

Let $H_n^{(h)}$ be the number of edges of $H_n^{(h)}$. Noting that in $H_n^{(h)}$ each non-empty independent $k$-subset covers exactly $k$ independent $(k-1)$-subsets, we can write

$$H_n^{(h)} = \sum_{k=1}^{\lceil n/(h+1)\rceil} k\, p_{n,k}^{(h)} = \sum_{k=1}^{\lceil n/(h+1)\rceil} k \binom{n-h\,k+h}{k}.$$

We immediately implement such formula to obtain a table of values of $H_n^{(h)}$.

```
H[h_, n_] := Sum[k p[h, n, k], {k, 1, Ceiling[n/(h + 1)]}]
```

| | n=0 | 1 | 2 | 3 | 4 | 5 | 6 | 7 | 8 | 9 | 10 | 11 |
|---|---|---|---|---|---|---|---|---|---|---|---|---|
| h=0 | 0 | 1 | 4 | 12 | 32 | 80 | 192 | 448 | 1024 | 2304 | 5120 | 11 264 |
| 1 | 0 | 1 | 2 | 5 | 10 | 20 | 38 | 71 | 130 | 235 | 420 | 744 |
| 2 | 0 | 1 | 2 | 3 | 6 | 11 | 18 | 30 | 50 | 81 | 130 | 208 |
| 3 | 0 | 1 | 2 | 3 | 4 | 7 | 12 | 19 | 28 | 42 | 64 | 97 |
| 4 | 0 | 1 | 2 | 3 | 4 | 5 | 8 | 13 | 20 | 29 | 40 | 56 |
| 5 | 0 | 1 | 2 | 3 | 4 | 5 | 6 | 9 | 14 | 21 | 30 | 41 |
| 6 | 0 | 1 | 2 | 3 | 4 | 5 | 6 | 7 | 10 | 15 | 22 | 31 |
| 7 | 0 | 1 | 2 | 3 | 4 | 5 | 6 | 7 | 8 | 11 | 16 | 23 |
| 8 | 0 | 1 | 2 | 3 | 4 | 5 | 6 | 7 | 8 | 9 | 12 | 17 |
| 9 | 0 | 1 | 2 | 3 | 4 | 5 | 6 | 7 | 8 | 9 | 10 | 13 |
| 10 | 0 | 1 | 2 | 3 | 4 | 5 | 6 | 7 | 8 | 9 | 10 | 11 |

## Remark.

For $h=1$, $H_n^{(h)}$ counts the number of edges of $\Gamma_n$.

Let now $T_{k,i}^{(n,h)}$ be the number of independent $k$-subsets of $P_n^{(h)}$ containing the vertex $v_i$, and let, for $h, k \geq 0$, $n \in \mathbb{Z}$,

$$\overline{p}_{n,k}^{(h)} = \begin{cases} p_{0,k}^{(h)} & \text{if } n < 0\,, \\ p_{n,k}^{(h)} & \text{if } n \geq 0\,. \end{cases}$$

```
pbar[h_, n_, k_] := If[n < 0, p[h, 0, k], p[h, n, k]]
```

## Lemma 3.1.

For $n, h, k \geq 0$, and $1 \leq i \leq n$,

$$T_{k,i}^{(n,h)} = \sum_{r=0}^{k-1} \overline{p}_{i-h-1,r}^{(h)}\, \overline{p}_{n-i-h,k-1-r}^{(h)}\,.$$

Next we show some tables of $T_{k,i}^{(n,h)}$.

```
T[h_, n_, k_, i_] := Sum[pbar[h, i - h - 1, r] pbar[h, n - i - h, k - 1 - r] , {r, 0, k - 1}]
```

## $T_{k,i}^{(7,2)}$

| | i=1 | 2 | 3 | 4 | 5 | 6 | 7 |
|---|---|---|---|---|---|---|---|
| k=0 | 0 | 0 | 0 | 0 | 0 | 0 | 0 |
| 1 | 1 | 1 | 1 | 1 | 1 | 1 | 1 |
| 2 | 4 | 3 | 2 | 2 | 2 | 3 | 4 |
| 3 | 1 | 0 | 0 | 1 | 0 | 0 | 1 |

## $T_{k,i}^{(10,2)}$

| | i=1 | 2 | 3 | 4 | 5 | 6 | 7 | 8 | 9 | 10 |
|---|---|---|---|---|---|---|---|---|---|---|
| k=0 | 0 | 0 | 0 | 0 | 0 | 0 | 0 | 0 | 0 | 0 |
| 1 | 1 | 1 | 1 | 1 | 1 | 1 | 1 | 1 | 1 | 1 |
| 2 | 7 | 6 | 5 | 5 | 5 | 5 | 5 | 5 | 6 | 7 |
| 3 | 10 | 6 | 3 | 5 | 6 | 6 | 5 | 3 | 6 | 10 |
| 4 | 1 | 0 | 0 | 1 | 0 | 0 | 1 | 0 | 0 | 1 |

## $T_{k,i}^{(15,3)}$

| | i=1 | 2 | 3 | 4 | 5 | 6 | 7 | 8 | 9 | 10 | 11 | 12 | 13 |
|---|---|---|---|---|---|---|---|---|---|---|---|---|---|
| k=0 | 0 | 0 | 0 | 0 | 0 | 0 | 0 | 0 | 0 | 0 | 0 | 0 | 0 |
| 1 | 1 | 1 | 1 | 1 | 1 | 1 | 1 | 1 | 1 | 1 | 1 | 1 | 1 |
| 2 | 11 | 10 | 9 | 8 | 8 | 8 | 8 | 8 | 8 | 8 | 8 | 8 | 9 |
| 3 | 28 | 21 | 15 | 10 | 13 | 15 | 16 | 16 | 16 | 15 | 13 | 10 | 15 |
| 4 | 10 | 4 | 1 | 0 | 6 | 6 | 3 | 0 | 3 | 6 | 6 | 0 | 1 |

## $T_{k,i}^{(6,1)}$

| | i=1 | 2 | 3 | 4 | 5 | 6 |
|---|---|---|---|---|---|---|
| k=0 | 0 | 0 | 0 | 0 | 0 | 0 |
| 1 | 1 | 1 | 1 | 1 | 1 | 1 |
| 2 | 4 | 3 | 3 | 3 | 3 | 4 |
| 3 | 3 | 1 | 2 | 2 | 1 | 3 |

## Remark.

$T_{k,i}^{(n,1)}$ counts the number of strings $\alpha = b_1 b_2 \cdots b_n \in \Gamma_n$ such that: (i) $H(\alpha, 00\cdots 0) = k$, and (ii) $b_i = 1$.

The following lemma provide a new formula for $H_n^{(h)}$.

## Lemma 3.2.

For positive $n$,

$$\sum_{k=1}^{\lceil n/(h+1)\rceil} \sum_{i=1}^{n} T_{k,i}^{(n,h)} = H_n^{(h)}.$$

Such new formula can be checked by implementing it in *Mathematica*, and comparing the values computed with the ones obtained above.

```
H2[h_, n_] := Sum[T[h, n, k, i], {k, 1, Ceiling[n/(h + 1)]}, {i, 1, n}]
```

| | n=0 | 1 | 2 | 3 | 4 | 5 | 6 | 7 | 8 | 9 | 10 | 11 |
|---|---|---|---|---|---|---|---|---|---|---|---|---|
| h=0 | 0 | 1 | 4 | 12 | 32 | 80 | 192 | 448 | 1024 | 2304 | 5120 | 11 264 |
| 1 | 0 | 1 | 2 | 5 | 10 | 20 | 38 | 71 | 130 | 235 | 420 | 744 |
| 2 | 0 | 1 | 2 | 3 | 6 | 11 | 18 | 30 | 50 | 81 | 130 | 208 |
| 3 | 0 | 1 | 2 | 3 | 4 | 7 | 12 | 19 | 28 | 42 | 64 | 97 |
| 4 | 0 | 1 | 2 | 3 | 4 | 5 | 8 | 13 | 20 | 29 | 40 | 56 |
| 5 | 0 | 1 | 2 | 3 | 4 | 5 | 6 | 9 | 14 | 21 | 30 | 41 |
| 6 | 0 | 1 | 2 | 3 | 4 | 5 | 6 | 7 | 10 | 15 | 22 | 31 |
| 7 | 0 | 1 | 2 | 3 | 4 | 5 | 6 | 7 | 8 | 11 | 16 | 23 |
| 8 | 0 | 1 | 2 | 3 | 4 | 5 | 6 | 7 | 8 | 9 | 12 | 17 |
| 9 | 0 | 1 | 2 | 3 | 4 | 5 | 6 | 7 | 8 | 9 | 10 | 13 |
| 10 | 0 | 1 | 2 | 3 | 4 | 5 | 6 | 7 | 8 | 9 | 10 | 11 |

Next we introduce a family of Fibonacci-like sequences.

## Definition 3.3.

For $h \geq 0$, and $n \geq 1$, we define the *h-Fibonacci sequence* $F^{(h)} = \{F_n^{(h)}\}_{n \geq 1}$ whose elements are

$$F_n^{(h)} = \begin{cases} 1 & \text{if } n \leq h+1, \\ F_{n-1}^{(h)} + F_{n-h-1}^{(h)} & \text{if } n > h+1. \end{cases}$$

The h-Fibonacci sequences are shown below, for $h = 0, \ldots, 10$.

```
F[h_, n_] := If[n ≤ h + 1, 1, F[h, n - 1] + F[h, n - h - 1]]
```

| | n=1 | 2 | 3 | 4 | 5 | 6 | 7 | 8 | 9 | 10 | 11 | 12 | 1 |
|---|---|---|---|---|---|---|---|---|---|---|---|---|---|
| h=0 | 1 | 2 | 4 | 8 | 16 | 32 | 64 | 128 | 256 | 512 | 1024 | 2048 | 40 |
| 1 | 1 | 1 | 2 | 3 | 5 | 8 | 13 | 21 | 34 | 55 | 89 | 144 | 23 |
| 2 | 1 | 1 | 1 | 2 | 3 | 4 | 6 | 9 | 13 | 19 | 28 | 41 | 6 |
| 3 | 1 | 1 | 1 | 1 | 2 | 3 | 4 | 5 | 7 | 10 | 14 | 19 | 2 |
| 4 | 1 | 1 | 1 | 1 | 1 | 2 | 3 | 4 | 5 | 6 | 8 | 11 | 1 |
| 5 | 1 | 1 | 1 | 1 | 1 | 1 | 2 | 3 | 4 | 5 | 6 | 7 | [illegible] |
| 6 | 1 | 1 | 1 | 1 | 1 | 1 | 1 | 2 | 3 | 4 | 5 | 6 | [illegible] |
| 7 | 1 | 1 | 1 | 1 | 1 | 1 | 1 | 1 | 2 | 3 | 4 | 5 | [illegible] |
| 8 | 1 | 1 | 1 | 1 | 1 | 1 | 1 | 1 | 1 | 2 | 3 | 4 | [illegible] |
| 9 | 1 | 1 | 1 | 1 | 1 | 1 | 1 | 1 | 1 | 1 | 2 | 3 | [illegible] |
| 10 | 1 | 1 | 1 | 1 | 1 | 1 | 1 | 1 | 1 | 1 | 1 | 2 | [illegible] |

From Lemma 2.2, and setting for $h \geq 0$, and $n \in \mathbb{Z}$,

$$\overline{p}_n^{(h)} = \begin{cases} p_0^{(h)} & \text{if } n < 0, \\ p_n^{(h)} & \text{if } n \geq 0, \end{cases}$$

we have that,

$$F_i^{(h)} = \overline{p}_{i-h-1}^{(h)}, \quad \text{for each } i \geq 1.$$

Thus, our Fibonacci-like sequences are obtained by adding a prefix of $h$ 1's to the sequence $p_0^{(h)}, p_1^{(h)}, \ldots$. Therefore, we have:

- $F^{(0)} = 1, 2, 4, \ldots, 2^n, \ldots$;
- $F^{(1)}$ is the Fibonacci sequence;
- more generally, $F^{(h)} = \underbrace{1, \ldots, 1}_{h}, p_0^{(h)}, p_1^{(h)}, p_2^{(h)}, \ldots$.

In the following, we use the discrete convolution operation $*$, as follows.

eq:convolution

$$\left(F^{(h)} * F^{(h)}\right)(n) \doteq \sum_{i=1}^{n} F_i^{(h)} F_{n-i+1}^{(h)}$$

## Theorem 3.4.

For $n$, $h \geq 0$, we have

$$H_n^{(h)} = \left(F^{(h)} * F^{(h)}\right)(n).$$

The previous theorem, our main result, provide a third way to compute the number of edges of $H_n^{(h)}$. Let us check again the correspondence between the tables obtained by the three different formulas presented in this section.

```
H3[h_, n_] := Sum[F[h, i] F[h, n - i + 1], {i, 1, n}]
```

| | n=0 | 1 | 2 | 3 | 4 | 5 | 6 | 7 | 8 | 9 | 10 | 11 |
|---|---|---|---|---|---|---|---|---|---|---|---|---|
| h=0 | 0 | 1 | 4 | 12 | 32 | 80 | 192 | 448 | 1024 | 2304 | 5120 | 11 264 |
| 1 | 0 | 1 | 2 | 5 | 10 | 20 | 38 | 71 | 130 | 235 | 420 | 744 |
| 2 | 0 | 1 | 2 | 3 | 6 | 11 | 18 | 30 | 50 | 81 | 130 | 208 |
| 3 | 0 | 1 | 2 | 3 | 4 | 7 | 12 | 19 | 28 | 42 | 64 | 97 |
| 4 | 0 | 1 | 2 | 3 | 4 | 5 | 8 | 13 | 20 | 29 | 40 | 56 |
| 5 | 0 | 1 | 2 | 3 | 4 | 5 | 6 | 9 | 14 | 21 | 30 | 41 |
| 6 | 0 | 1 | 2 | 3 | 4 | 5 | 6 | 7 | 10 | 15 | 22 | 31 |
| 7 | 0 | 1 | 2 | 3 | 4 | 5 | 6 | 7 | 8 | 11 | 16 | 23 |
| 8 | 0 | 1 | 2 | 3 | 4 | 5 | 6 | 7 | 8 | 9 | 12 | 17 |
| 9 | 0 | 1 | 2 | 3 | 4 | 5 | 6 | 7 | 8 | 9 | 10 | 13 |
| 10 | 0 | 1 | 2 | 3 | 4 | 5 | 6 | 7 | 8 | 9 | 10 | 11 |

## Remark.

For $h = 1$, we obtain the number of edges of $\Gamma_n$ by using Fibonacci numbers:

$$H_n^{(h)} = \sum_{i=1}^{n} F_i\, F_{n-i+1}.$$

The latter result is [Kla05, Proposition 3].

# 4. The independent subsets of powers of cycles

For $n$, $h$, $k \geq 0$, we denote by $q_{n,k}^{(h)}$ the number of independent $k$-subsets of $Q_n^{(h)}$.

## Remark.

For $h = 1$, $n > 1$, $q_{n,k}^{(h)}$ counts the number of binary strings $\alpha \in \Lambda_n$ such that $H(\alpha, 00 \cdots 0) = k$.

## Lemma 4.1.

For $n$, $h \geq 0$, and $k > 1$,

$$q_{n,k}^{(h)} = \frac{n}{k}\binom{n - h\,k - 1}{k - 1}.$$

We compute some values of $q_{n,k}^{(h)}$, for $h = 1, \ldots, 4$.

```
q[h_, n_, k_] := n/k MyBinomial[n - h k - 1, k - 1]; q[h_, n_, 0] := 1; q[h_, n_, 1] := n
```

## $q_{n,k}{}^{(1)}$

| | n=0 | 1 | 2 | 3 | 4 | 5 | 6 | 7 | 8 | 9 | 10 | 11 | 12 | 13 | 14 | 15 | 16 |
|---|---|---|---|---|---|---|---|---|---|---|---|---|---|---|---|---|---|
| k=0 | 1 | 1 | 1 | 1 | 1 | 1 | 1 | 1 | 1 | 1 | 1 | 1 | 1 | 1 | 1 | 1 | 1 |
| 1 | 0 | 1 | 2 | 3 | 4 | 5 | 6 | 7 | 8 | 9 | 10 | 11 | 12 | 13 | 14 | 15 | 16 |
| 2 | 0 | 0 | 0 | 0 | 2 | 5 | 9 | 14 | 20 | 27 | 35 | 44 | 54 | 65 | 77 | 90 | 104 |
| 3 | 0 | 0 | 0 | 0 | 0 | 0 | 2 | 7 | 16 | 30 | 50 | 77 | 112 | 156 | 210 | 275 | 352 |
| 4 | 0 | 0 | 0 | 0 | 0 | 0 | 0 | 0 | 2 | 9 | 25 | 55 | 105 | 182 | 294 | 450 | 660 |
| 5 | 0 | 0 | 0 | 0 | 0 | 0 | 0 | 0 | 0 | 0 | 2 | 11 | 36 | 91 | 196 | 378 | 672 |
| 6 | 0 | 0 | 0 | 0 | 0 | 0 | 0 | 0 | 0 | 0 | 0 | 0 | 2 | 13 | 49 | 140 | 336 |
| 7 | 0 | 0 | 0 | 0 | 0 | 0 | 0 | 0 | 0 | 0 | 0 | 0 | 0 | 0 | 2 | 15 | 64 |
| 8 | 0 | 0 | 0 | 0 | 0 | 0 | 0 | 0 | 0 | 0 | 0 | 0 | 0 | 0 | 0 | 0 | 2 |

## $q_{n,k}{}^{(2)}$

| | n=0 | 1 | 2 | 3 | 4 | 5 | 6 | 7 | 8 | 9 | 10 | 11 | 12 | 13 | 14 | 15 | 16 | 17 |
|---|---|---|---|---|---|---|---|---|---|---|---|---|---|---|---|---|---|---|
| k=0 | 1 | 1 | 1 | 1 | 1 | 1 | 1 | 1 | 1 | 1 | 1 | 1 | 1 | 1 | 1 | 1 | 1 | 1 |
| 1 | 0 | 1 | 2 | 3 | 4 | 5 | 6 | 7 | 8 | 9 | 10 | 11 | 12 | 13 | 14 | 15 | 16 | 17 |
| 2 | 0 | 0 | 0 | 0 | 0 | 0 | 3 | 7 | 12 | 18 | 25 | 33 | 42 | 52 | 63 | 75 | 88 | 102 |
| 3 | 0 | 0 | 0 | 0 | 0 | 0 | 0 | 0 | 0 | 3 | 10 | 22 | 40 | 65 | 98 | 140 | 192 | 255 |
| 4 | 0 | 0 | 0 | 0 | 0 | 0 | 0 | 0 | 0 | 0 | 0 | 0 | 3 | 13 | 35 | 75 | 140 | 238 |
| 5 | 0 | 0 | 0 | 0 | 0 | 0 | 0 | 0 | 0 | 0 | 0 | 0 | 0 | 0 | 0 | 3 | 16 | 51 |

## $q_{n,k}{}^{(3)}$

| | n=0 | 1 | 2 | 3 | 4 | 5 | 6 | 7 | 8 | 9 | 10 | 11 | 12 | 13 | 14 |
|---|---|---|---|---|---|---|---|---|---|---|---|---|---|---|---|
| k=0 | 1 | 1 | 1 | 1 | 1 | 1 | 1 | 1 | 1 | 1 | 1 | 1 | 1 | 1 | 1 |
| 1 | 0 | 1 | 2 | 3 | 4 | 5 | 6 | 7 | 8 | 9 | 10 | 11 | 12 | 13 | 14 |
| 2 | 0 | 0 | 0 | 0 | 0 | 0 | 0 | 0 | 4 | 9 | 15 | 22 | 30 | 39 | 49 |
| 3 | 0 | 0 | 0 | 0 | 0 | 0 | 0 | 0 | 0 | 0 | 0 | 0 | 4 | 13 | 28 |
| 4 | 0 | 0 | 0 | 0 | 0 | 0 | 0 | 0 | 0 | 0 | 0 | 0 | 0 | 0 | 0 |

## $q_{n,k}{}^{(4)}$

| | n=0 | 1 | 2 | 3 | 4 | 5 | 6 | 7 | 8 | 9 | 10 | 11 | 12 | 13 | 14 |
|---|---|---|---|---|---|---|---|---|---|---|---|---|---|---|---|
| k=0 | 1 | 1 | 1 | 1 | 1 | 1 | 1 | 1 | 1 | 1 | 1 | 1 | 1 | 1 | 1 |
| 1 | 0 | 1 | 2 | 3 | 4 | 5 | 6 | 7 | 8 | 9 | 10 | 11 | 12 | 13 | 14 |
| 2 | 0 | 0 | 0 | 0 | 0 | 0 | 0 | 0 | 0 | 0 | 5 | 11 | 18 | 26 | 35 |
| 3 | 0 | 0 | 0 | 0 | 0 | 0 | 0 | 0 | 0 | 0 | 0 | 0 | 0 | 0 | 0 |

Moreover, $q_{n,0}^{(h)} = 1$, and $q_{n,1}^{(h)} = n$, for each $n$, $h \geq 0$.

For $n$, $h \geq 0$, the number of all independent subsets of $Q_n^{(h)}$ is

$$q_n^{(h)} = \sum_{k \geq 0} q_{n,k}^{(h)} = \sum_{k=0}^{\lceil n/(h+1) \rceil} q_{n,k}^{(h)},$$

The formula for the number of all independent subsets of $Q_n^{(h)}$ is implemented below.

```
q[h_, n_] := Sum[q[h, n, k], {k, 0, Ceiling[n/(h + 1)]}]
```

| | n=0 | 1 | 2 | 3 | 4 | 5 | 6 | 7 | 8 | 9 | 10 | 11 | 12 | 13 | 14 | 15 | 16 | |
|---|---|---|---|---|---|---|---|---|---|---|---|---|---|---|---|---|---|---|
| h=0 | 1 | 2 | 4 | 8 | 16 | 32 | 64 | 128 | 256 | 512 | 1024 | 2048 | 4096 | 8192 | 16384 | 32768 | 65536 | 13 |
| 1 | 1 | 2 | 3 | 4 | 7 | 11 | 18 | 29 | 47 | 76 | 123 | 199 | 322 | 521 | 843 | 1364 | 2207 | 3 |
| 2 | 1 | 2 | 3 | 4 | 5 | 6 | 10 | 15 | 21 | 31 | 46 | 67 | 98 | 144 | 211 | 309 | 453 | ( |
| 3 | 1 | 2 | 3 | 4 | 5 | 6 | 7 | 8 | 13 | 19 | 26 | 34 | 47 | 66 | 92 | 126 | 173 | : |
| 4 | 1 | 2 | 3 | 4 | 5 | 6 | 7 | 8 | 9 | 10 | 16 | 23 | 31 | 40 | 50 | 66 | 89 | : |
| 5 | 1 | 2 | 3 | 4 | 5 | 6 | 7 | 8 | 9 | 10 | 11 | 12 | 19 | 27 | 36 | 46 | 57 | |
| 6 | 1 | 2 | 3 | 4 | 5 | 6 | 7 | 8 | 9 | 10 | 11 | 12 | 13 | 14 | 22 | 31 | 41 | |
| 7 | 1 | 2 | 3 | 4 | 5 | 6 | 7 | 8 | 9 | 10 | 11 | 12 | 13 | 14 | 15 | 16 | 25 | |
| 8 | 1 | 2 | 3 | 4 | 5 | 6 | 7 | 8 | 9 | 10 | 11 | 12 | 13 | 14 | 15 | 16 | 17 | |
| 9 | 1 | 2 | 3 | 4 | 5 | 6 | 7 | 8 | 9 | 10 | 11 | 12 | 13 | 14 | 15 | 16 | 17 | |
| 10 | 1 | 2 | 3 | 4 | 5 | 6 | 7 | 8 | 9 | 10 | 11 | 12 | 13 | 14 | 15 | 16 | 17 | |

## Remark.

Denote by $L_n$ the $n^{th}$ element of the Lucas sequence $L_1 = 1$, $L_2 = 3$, and $L_i = L_{i-1} + L_{i-2}$, for $i > 2$ (some values of the Lucas sequence are shown below). Then, for $n > 1$, $q_n^{(1)} = L_n$ is the number of elements of the Lucas cube of order $n$.

```
MyLucas[n_] := MyLucas[n - 1] + MyLucas[n - 2]; MyLucas[1] := 1; MyLucas[2] := 3
```

| n=1 | 2 | 3 | 4 | 5 | 6 | 7 | 8 | 9 | 10 | 11 | 12 | 13 |
|---|---|---|---|---|---|---|---|---|---|---|---|---|
| 1 | 3 | 4 | 7 | 11 | 18 | 29 | 47 | 76 | 123 | 199 | 322 | 521 |

The coefficients $q_n^{(h)}$ satisfy a recursion that closely resemble that of Lemma 2.2.

## Lemma 4.2.

For $n$, $h \geq 0$,

eq:q_n recurrence

$$q_n^{(h)} = \begin{cases} n+1 & \text{if } n \leq 2h+1, \\ q_{n-1}^{(h)} + q_{n-h-1}^{(h)} & \text{if } n > 2h+1. \end{cases}$$

We check the two provided formulas for $q_n^{(h)}$ indeed compute the same numbers (at least for the values appearing in the tables).

```
q2[h_, n_] := If[n > 2 h + 1, q[h, n - 1] + q[h, n - h - 1], n + 1]
```

| | n=0 | 1 | 2 | 3 | 4 | 5 | 6 | 7 | 8 | 9 | 10 | 11 | 12 | 13 | 14 | 15 | 16 | |
|---|---|---|---|---|---|---|---|---|---|---|---|---|---|---|---|---|---|---|
| h=0 | 1 | 2 | 4 | 8 | 16 | 32 | 64 | 128 | 256 | 512 | 1024 | 2048 | 4096 | 8192 | 16384 | 32768 | 65536 | 13 |
| 1 | 1 | 2 | 3 | 4 | 7 | 11 | 18 | 29 | 47 | 76 | 123 | 199 | 322 | 521 | 843 | 1364 | 2207 | 3 |
| 2 | 1 | 2 | 3 | 4 | 5 | 6 | 10 | 15 | 21 | 31 | 46 | 67 | 98 | 144 | 211 | 309 | 453 | ( |
| 3 | 1 | 2 | 3 | 4 | 5 | 6 | 7 | 8 | 13 | 19 | 26 | 34 | 47 | 66 | 92 | 126 | 173 | : |
| 4 | 1 | 2 | 3 | 4 | 5 | 6 | 7 | 8 | 9 | 10 | 16 | 23 | 31 | 40 | 50 | 66 | 89 | : |
| 5 | 1 | 2 | 3 | 4 | 5 | 6 | 7 | 8 | 9 | 10 | 11 | 12 | 19 | 27 | 36 | 46 | 57 | |
| 6 | 1 | 2 | 3 | 4 | 5 | 6 | 7 | 8 | 9 | 10 | 11 | 12 | 13 | 14 | 22 | 31 | 41 | |
| 7 | 1 | 2 | 3 | 4 | 5 | 6 | 7 | 8 | 9 | 10 | 11 | 12 | 13 | 14 | 15 | 16 | 25 | |
| 8 | 1 | 2 | 3 | 4 | 5 | 6 | 7 | 8 | 9 | 10 | 11 | 12 | 13 | 14 | 15 | 16 | 17 | |
| 9 | 1 | 2 | 3 | 4 | 5 | 6 | 7 | 8 | 9 | 10 | 11 | 12 | 13 | 14 | 15 | 16 | 17 | |
| 10 | 1 | 2 | 3 | 4 | 5 | 6 | 7 | 8 | 9 | 10 | 11 | 12 | 13 | 14 | 15 | 16 | 17 | |

# 5. The poset of independent subsets of powers of cycles

Next figure  shows a few Hasse diagrams $M_n^{(h)}$. Notice that, as stated in the introduction, for each $n$,

$M_n^{(1)}$ is the Lucas cube $\Lambda_n$.

Out[279]=

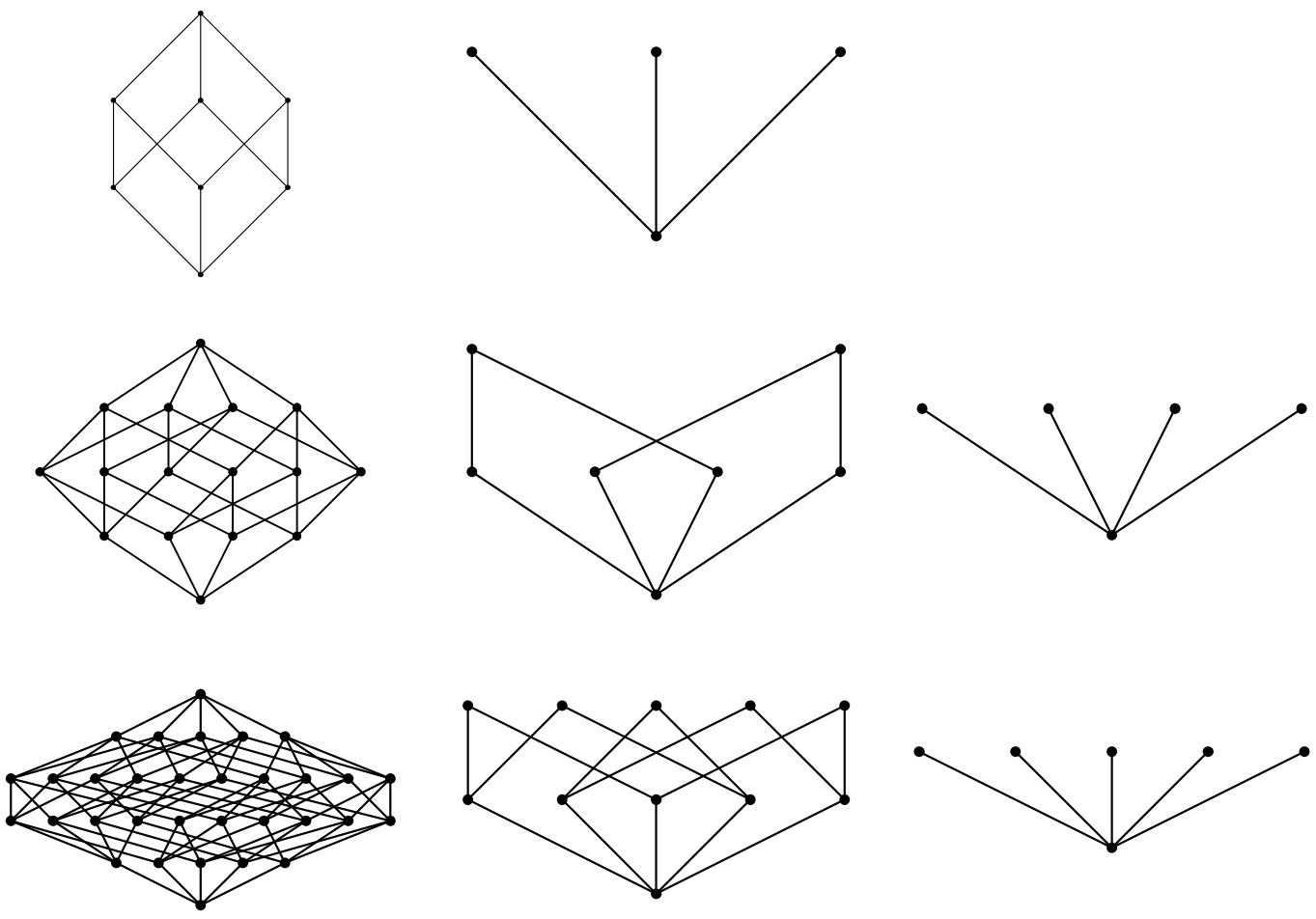

Let $M_n^{(h)}$ be the number of edges of $M_n^{(h)}$. As done in Section 3 for the case of paths, we immediately provide a formula for $M_n^{(h)}$:

eq:M_nh formula

$$M_n^{(h)} = \sum_{k=0}^{\lceil n/h+1 \rceil} k\, q_{n,k}^{(h)} = n \sum_{k=0}^{\lceil n/h+1 \rceil} \binom{n-h\,k-1}{k-1}.$$

```
M[h_, n_] := Sum[k q[h, n, k], {k, 1, Ceiling[n/(h + 1)]}]
```

| | n=0 | 1 | 2 | 3 | 4 | 5 | 6 | 7 | 8 | 9 | 10 | 11 | 12 | 13 | 14 | 15 |
|---|---|---|---|---|---|---|---|---|---|---|---|---|---|---|---|---|
| h=0 | 0 | 1 | 4 | 12 | 32 | 80 | 192 | 448 | 1024 | 2304 | 5120 | 11 264 | 24 576 | 53 248 | 114 688 | 245 760 |
| 1 | 0 | 1 | 2 | 3 | 8 | 15 | 30 | 56 | 104 | 189 | 340 | 605 | 1068 | 1872 | 3262 | 5655 |
| 2 | 0 | 1 | 2 | 3 | 4 | 5 | 12 | 21 | 32 | 54 | 90 | 143 | 228 | 364 | 574 | 900 |
| 3 | 0 | 1 | 2 | 3 | 4 | 5 | 6 | 7 | 16 | 27 | 40 | 55 | 84 | 130 | 196 | 285 |
| 4 | 0 | 1 | 2 | 3 | 4 | 5 | 6 | 7 | 8 | 9 | 20 | 33 | 48 | 65 | 84 | 120 |
| 5 | 0 | 1 | 2 | 3 | 4 | 5 | 6 | 7 | 8 | 9 | 10 | 11 | 24 | 39 | 56 | 75 |
| 6 | 0 | 1 | 2 | 3 | 4 | 5 | 6 | 7 | 8 | 9 | 10 | 11 | 12 | 13 | 28 | 45 |
| 7 | 0 | 1 | 2 | 3 | 4 | 5 | 6 | 7 | 8 | 9 | 10 | 11 | 12 | 13 | 14 | 15 |
| 8 | 0 | 1 | 2 | 3 | 4 | 5 | 6 | 7 | 8 | 9 | 10 | 11 | 12 | 13 | 14 | 15 |
| 9 | 0 | 1 | 2 | 3 | 4 | 5 | 6 | 7 | 8 | 9 | 10 | 11 | 12 | 13 | 14 | 15 |
| 10 | 0 | 1 | 2 | 3 | 4 | 5 | 6 | 7 | 8 | 9 | 10 | 11 | 12 | 13 | 14 | 15 |

## Remark.

For $h = 1$, $n > 1$, $M_n^{(h)}$ counts the number of edges of $\Lambda_n$. As shown in [MPCZS01, Proposition 4(ii)], $M_n^{(h)} = n\,F_{n-1}$.

```
TableForm[Table[n MyFibo[n - 1], {n, 2, 15}],
 TableHeadings → {{"n=2", "3", "4", "5", "6", "7", "8", "9", "10",
    "11", "12", "13", "14", "15", "16", "17", "18", "19", "20"}},
 TableAlignments → Center, TableDirections → Row]
```

| n=2 | 3 | 4 | 5 | 6 | 7 | 8 | 9 | 10 | 11 | 12 | 13 | 14 | 15 |
|---|---|---|---|---|---|---|---|---|---|---|---|---|---|
| 2 | 3 | 8 | 15 | 30 | 56 | 104 | 189 | 340 | 605 | 1068 | 1872 | 3262 | 5655 |

As shown in the proof of Lemma 4.1 (see [CD12a]), the value

$$p_{n-2\,h-1,k-1}^{(h)} = \binom{n-h\,k-1}{k-1}$$

is the analogue of the coefficient $T_{k,i}^{(n,h)}$: in the case of cycles we have no dependencies on $i$,

because each choice of vertex is equivalent. We can obtain $M_n^{(h)}$ in terms of a fibonacci-like sequence, as follows.

## Proposition 5.1.

For $n > h \geq 0$, the following holds.

$$M_n^{(h)} = n\,F_{n-h}^{(h)}.$$

```
M2[h_, n_] := n F[h, n - h]
```

| | n=0 | 1 | 2 | 3 | 4 | 5 | 6 | 7 | 8 | 9 | 10 | 11 | 12 | 13 | 14 | 15 |
|---|---|---|---|---|---|---|---|---|---|---|---|---|---|---|---|---|
| h=0 | 0 | 1 | 4 | 12 | 32 | 80 | 192 | 448 | 1024 | 2304 | 5120 | 11 264 | 24 576 | 53 248 | 114 688 | 245 760 |
| 1 | 0 | 1 | 2 | 3 | 8 | 15 | 30 | 56 | 104 | 189 | 340 | 605 | 1068 | 1872 | 3262 | 5655 |
| 2 | 0 | 1 | 2 | 3 | 4 | 5 | 12 | 21 | 32 | 54 | 90 | 143 | 228 | 364 | 574 | 900 |
| 3 | 0 | 1 | 2 | 3 | 4 | 5 | 6 | 7 | 16 | 27 | 40 | 55 | 84 | 130 | 196 | 285 |
| 4 | 0 | 1 | 2 | 3 | 4 | 5 | 6 | 7 | 8 | 9 | 20 | 33 | 48 | 65 | 84 | 120 |
| 5 | 0 | 1 | 2 | 3 | 4 | 5 | 6 | 7 | 8 | 9 | 10 | 11 | 24 | 39 | 56 | 75 |
| 6 | 0 | 1 | 2 | 3 | 4 | 5 | 6 | 7 | 8 | 9 | 10 | 11 | 12 | 13 | 28 | 45 |
| 7 | 0 | 1 | 2 | 3 | 4 | 5 | 6 | 7 | 8 | 9 | 10 | 11 | 12 | 13 | 14 | 15 |
| 8 | 0 | 1 | 2 | 3 | 4 | 5 | 6 | 7 | 8 | 9 | 10 | 11 | 12 | 13 | 14 | 15 |
| 9 | 0 | 1 | 2 | 3 | 4 | 5 | 6 | 7 | 8 | 9 | 10 | 11 | 12 | 13 | 14 | 15 |
| 10 | 0 | 1 | 2 | 3 | 4 | 5 | 6 | 7 | 8 | 9 | 10 | 11 | 12 | 13 | 14 | 15 |

# Toward the analogous of Theorem 3.4.

In [CD12a] we have not provided an analogous of the Theorem 3.4 for the case of cycles. In order to obtain such result, the first step is to generalize Lucas numbers in an appropriate way. After several attempts, we can assume that the generalization of the Lucas numbers which agrees with our combinatorial structures is the following.

```
L[h_, n_] := If[n ≤ h + 1, 1, L[h, n - 1] + L[h, n - h - 1]]; L[h_, 1] := h + 1
```

| | n=1 | 2 | 3 | 4 | 5 | 6 | 7 | 8 | 9 | 10 | 11 | 12 | 13 | 14 | 15 |
|---|---|---|---|---|---|---|---|---|---|---|---|---|---|---|---|
| h=0 | 1 | 2 | 4 | 8 | 16 | 32 | 64 | 128 | 256 | 512 | 1024 | 2048 | 4096 | 8192 | 16 384 |
| 1 | 2 | 1 | 3 | 4 | 7 | 11 | 18 | 29 | 47 | 76 | 123 | 199 | 322 | 521 | 843 |
| 2 | 3 | 1 | 1 | 4 | 5 | 6 | 10 | 15 | 21 | 31 | 46 | 67 | 98 | 144 | 211 |
| 3 | 4 | 1 | 1 | 1 | 5 | 6 | 7 | 8 | 13 | 19 | 26 | 34 | 47 | 66 | 92 |
| 4 | 5 | 1 | 1 | 1 | 1 | 6 | 7 | 8 | 9 | 10 | 16 | 23 | 31 | 40 | 50 |
| 5 | 6 | 1 | 1 | 1 | 1 | 1 | 7 | 8 | 9 | 10 | 11 | 12 | 19 | 27 | 36 |
| 6 | 7 | 1 | 1 | 1 | 1 | 1 | 1 | 8 | 9 | 10 | 11 | 12 | 13 | 14 | 22 |
| 7 | 8 | 1 | 1 | 1 | 1 | 1 | 1 | 1 | 9 | 10 | 11 | 12 | 13 | 14 | 15 |
| 8 | 9 | 1 | 1 | 1 | 1 | 1 | 1 | 1 | 1 | 10 | 11 | 12 | 13 | 14 | 15 |
| 9 | 10 | 1 | 1 | 1 | 1 | 1 | 1 | 1 | 1 | 1 | 11 | 12 | 13 | 14 | 15 |
| 10 | 11 | 1 | 1 | 1 | 1 | 1 | 1 | 1 | 1 | 1 | 1 | 12 | 13 | 14 | 15 |

Following the terminology adopted in Section 3, we call this sequences the *h-Lucas sequences*, and denote them by $L^{(h)} = \{L_n^{(h)}\}_{n\geq 1}$. Notice that $L_{n+1}^{(1)} = L_n$ for each $n \geq 1$.

The following function seems to be able to generate, although only for the case $n \geq h$, our values $M_n^{(h)}$, by using an appropriate discrete convolution of an $h$-Lucas sequence and an $h$-Fibonacci sequence. The latter fact can be checked by comparing the following table with the table of values of $M_n^{(h)}$ provided in the beginning of this section.

```
Mtent[h_, n_] := Sum[F[h, i] L[h, n - h - i + 1], {i, 1, n - h}]
```

```
TableForm[Table[Mtent[h, n], {h, 0, 10}, {n, 0, 15}],
 TableHeadings → {{"h=0", "1", "2", "3", "4", "5", "6", "7", "8", "9", "10"},
   {"n=0", "1", "2", "3", "4", "5", "6", "7", "8", "9", "10", "11", "12", "13",
    "14", "15", "16", "17", "18", "19", "20"}}, TableAlignments → Center]
```

| | n=0 | 1 | 2 | 3 | 4 | 5 | 6 | 7 | 8 | 9 | 10 | 11 | 12 | 13 | 14 | 15 |
|---|---|---|---|---|---|---|---|---|---|---|---|---|---|---|---|---|
| h=0 | 0 | 1 | 4 | 12 | 32 | 80 | 192 | 448 | 1024 | 2304 | 5120 | 11 264 | 24 576 | 53 248 | 114 688 | 245 760 |
| 1 | 0 | 0 | 2 | 3 | 8 | 15 | 30 | 56 | 104 | 189 | 340 | 605 | 1068 | 1872 | 3262 | 5655 |
| 2 | 0 | 0 | 0 | 3 | 4 | 5 | 12 | 21 | 32 | 54 | 90 | 143 | 228 | 364 | 574 | 900 |
| 3 | 0 | 0 | 0 | 0 | 4 | 5 | 6 | 7 | 16 | 27 | 40 | 55 | 84 | 130 | 196 | 285 |
| 4 | 0 | 0 | 0 | 0 | 0 | 5 | 6 | 7 | 8 | 9 | 20 | 33 | 48 | 65 | 84 | 120 |
| 5 | 0 | 0 | 0 | 0 | 0 | 0 | 6 | 7 | 8 | 9 | 10 | 11 | 24 | 39 | 56 | 75 |
| 6 | 0 | 0 | 0 | 0 | 0 | 0 | 0 | 7 | 8 | 9 | 10 | 11 | 12 | 13 | 28 | 45 |
| 7 | 0 | 0 | 0 | 0 | 0 | 0 | 0 | 0 | 8 | 9 | 10 | 11 | 12 | 13 | 14 | 15 |
| 8 | 0 | 0 | 0 | 0 | 0 | 0 | 0 | 0 | 0 | 9 | 10 | 11 | 12 | 13 | 14 | 15 |
| 9 | 0 | 0 | 0 | 0 | 0 | 0 | 0 | 0 | 0 | 0 | 10 | 11 | 12 | 13 | 14 | 15 |
| 10 | 0 | 0 | 0 | 0 | 0 | 0 | 0 | 0 | 0 | 0 | 0 | 11 | 12 | 13 | 14 | 15 |

Thanks to the aid of *Mathematica*, we can thus conjecture the desidered analogous of Theorem 3.4, as follows.

### Conjecture 5.2.

For $n > h \geq 0$, we have

$$M_n^{(h)} = \left(F^{(h)} * L^{(h)}\right)(n - h).$$